\documentclass[twocolumn,english,reprint,longbibliography,prl]{revtex4-1}
\usepackage[utf8]{inputenc}
\usepackage{amsmath}
\usepackage{xcolor}
\usepackage{graphicx}
\usepackage{dcolumn}
\usepackage{bm}
\usepackage{comment}
\usepackage[normalem]{ulem}
\usepackage{lipsum}
\usepackage{color}
\usepackage{xcolor}
\usepackage[normalem]{ulem}
\usepackage{simplewick}
\usepackage{qcircuit}
\usepackage{braket}
\usepackage{float}
\usepackage{multirow}
\usepackage{tikz}
\usepackage[utf8]{inputenc}
\usepackage{multibib}
\usepackage{appendix}
\usepackage[colorlinks=true,%
bookmarks=false,%
linkcolor=blue,%
urlcolor=blue,%
citecolor=blue,%
breaklinks]{hyperref}
\usepackage{soul}

\usepackage{braket}
\usepackage{subfig}

\makeatother

\begin{document}

\author{Guglielmo Mazzola}
\affiliation{Institute for Computational Science, University of Zurich, Winterthurerstrasse 190, 8057 Zurich, Switzerland}

\author{Giuseppe~Carleo}
\affiliation{Institute of Physics, École Polytechnique Fédérale de Lausanne (EPFL), CH-1015 Lausanne, Switzerland}

\date{today}

\title{Exponential challenges in unbiasing quantum Monte Carlo algorithms with quantum computers }
\date{\today}

\begin{abstract}

Recently, Huggins et.~al. \cite{huggins2022unbiasing} devised a general  projective Quantum Monte Carlo method suitable for implementation on quantum computers.
This hybrid approach, however, relies on a subroutine --the computation of the local energy estimator on the quantum computer-- that is intrinsically affected by an exponential scaling of the computational time with the number of qubits. By means of numerical experiments, we show that this exponential scaling manifests prominently already on systems below the point of ``quantum advantage". For the prototypical transverse-field Ising model, we show that the required time resources to compete with classical simulations on around 40 qubits are already of the order of $10^{13}$ projective measurements, with an estimated running time of a few thousand years on superconducting hardware. These observations strongly suggest that the proposed hybrid method, in its present form, is unlikely to offer a sizeable advantage over conventional quantum Monte Carlo approaches.

\end{abstract}

\maketitle

Quantum Monte Carlo (QMC) methods represent invaluable tools to simulate many-body quantum systems on classical (henceforth also named conventional) computers, including models of interacting spins, bosons, fermions, and electrons in real space \cite{ceperley_ground_1980}.
Despite the power of QMC methods, there exist many important cases that are not efficiently tractable without approximations, because of the infamous \emph{sign-problem} \cite{troyer_computational_2005}. This family of classically intractable systems, including frustrated spins and electronic problems, is a natural candidate for quantum speed up on quantum computers.
Recently, Huggins et. al.\cite{huggins2022unbiasing} devised a general implementation of projective QMC methods for quantum computers, henceforth named QC-QMC, combining one of the most powerful class of classical algorithms to understand interacting quantum particles, with disrupting new possibilities offered by quantum computing hardware. The algorithm is generally referred to as being ``scalable and noise resilient", while potential issues related to its asymptotic scaling and the possibility of quantum advantage are discussed in Supplementary Section F \cite{huggins2022unbiasing}.
In this note we clarify that the QC-QMC approach  is intrinsically affected by an exponential scaling of the computational time with the number of qubits, because of the exponential scaling of the number of measurements required to estimate the so-called ``local energy". By means of numerical experiments, we show that this exponential scaling manifests prominently already on systems below the point of "quantum advantage", and is particularly severe already on a simple sign-problem free spin Hamiltonian. These findings strongly suggest that it is unlikely that QC-QMC can offer, in its present form, a realistic improvement over pure QMC approaches.

To elucidate challenges in porting a conventional QMC algorithm to a quantum computing setting we focus on analyzing a key subroutine of the QC-QMC method. Specifically, we focus on the cost of computing the ``local energy" estimator, a common bulding block of the vast majority of QMC algorithms \cite{foulkes_quantum_2001,Becca2017}. For the sake of specifing a concrete problem, we choose the textbook transvere-field Ising (TFI) model, encapsulating salient features of strong correlated quantum physics.
At variance with the quantum chemistry models of Ref.~\cite{huggins2022unbiasing}, the TFI model is a simpler setup for numerical experiments, and it is also \emph{sign-problem} free, thus representing a controlled and well studied benchmark against exact solutions obtained by means of conventional QMC methods.


We define the ferromagnetic TFI hamiltonian, on a one-dimensional periodic $L$-spin system as
\begin{equation}
\label{eq:ham}
    H = - J \sum_{k}^L \sigma^z_k \sigma^z_{k+1} - \Gamma \sum_{k}^L \sigma^x_k ,
\end{equation}
where $\sigma^\alpha$ are Pauli matrices,
and consider in the following numerical simulations the critical transition point at $J = \Gamma =1$.
We denote a generic computational basis configuration as $|x\rangle = (s_1,\cdots, s_L)$, where $s_k$ are eigenvalues $\{1,-1\}$ of the $\sigma_j^z$ operator. 
Each basis vector can be conveniently labeled by its integer representation, by assuming the standard spin-$1/2$ to qubit mapping, $1\rightarrow 0$ and $-1\rightarrow 1$.
Given a configuration $|x\rangle$, we denote its connected elements by $|x'\rangle$. These are the $L$ configurations for which the matrix element
\begin{equation}
    H_{x,x'} = \langle x' | H | x \rangle
\end{equation}
is non zero, $|H_{x,x'}| \neq 0$. Namely, this set includes the $L$ configurations that are connected by a single spin-flip to $x$, plus $x$ itself.
In the AFQMC case discussed in Ref~\cite{huggins2022unbiasing}, the walker $x$ are fermionic occupation numbers in a given second-quantized representation.

\begin{figure*}[hbt!]
\includegraphics[width=0.90\textwidth]{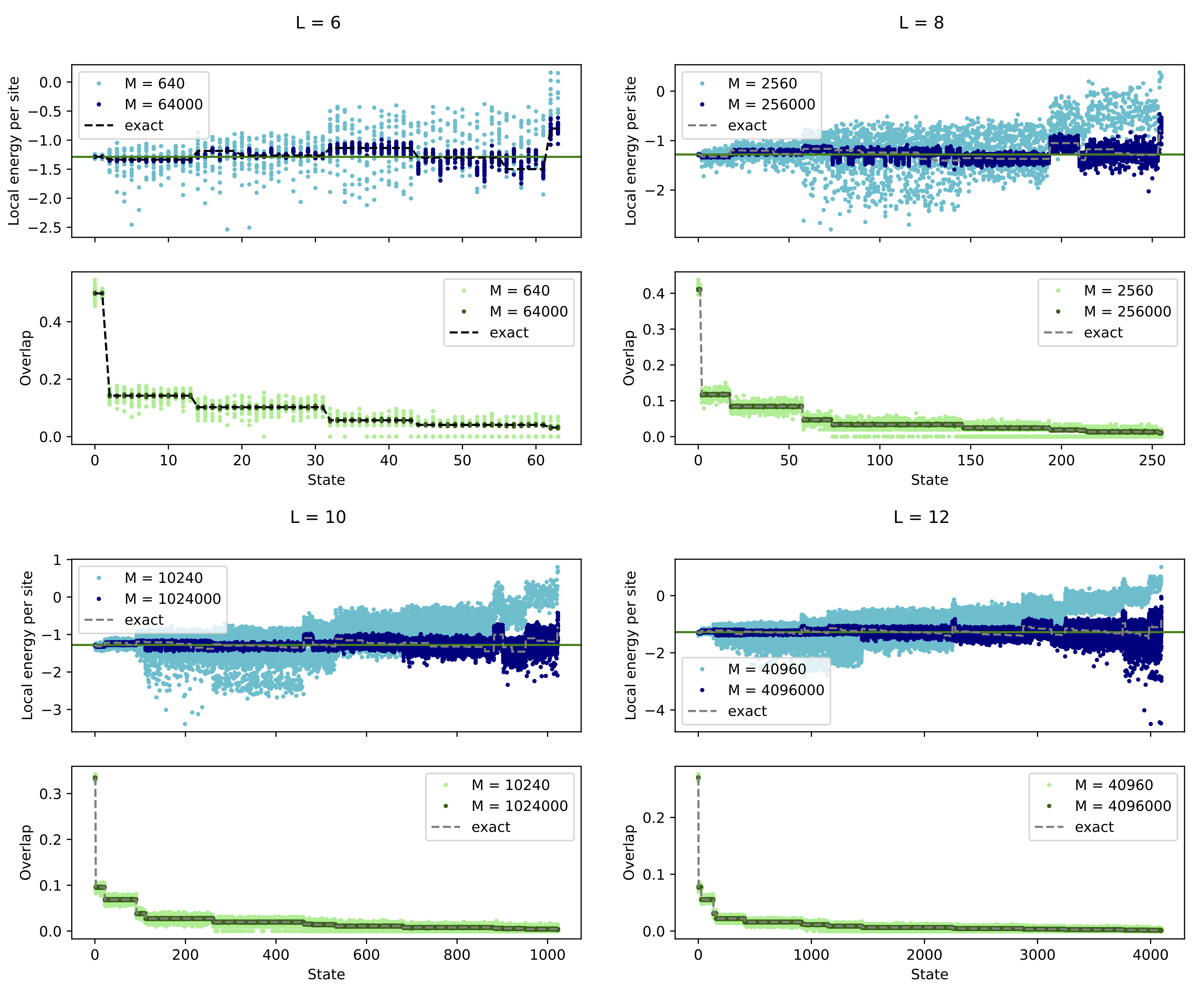}
\caption{
Statistical fluctuations of the local energy $e_L(x)$ (above) and the overlaps (below) for all states $x$ (sorted by decreasing overlap magnitude) as a function of $M$, for different system sizes, $L=6,8,10$, and $12$. For each size, we emulate the quantum measurement process using $M = M_0 \cdot 2^L$, with $M_0=10,1000$ shots. The statistical error in the local energy is of order $100\%$ for an exponentially increasing number of states, $x$, even if we also increase exponentially, with $L$ the number of shots. The dashed lines correspond to the exact values given by a conventional, or noiseless, evaluation using $\psi_T(x)$. The continuous green line denotes the ground state energy.
}
\label{fig:le}
\end{figure*}

The exact ground state of Eq.~\ref{eq:ham}, $|\psi_0(x)\rangle$ is a positive definite linear combination of $2^L$ components.
We also denote by $\psi_T(x)$ a trial wavefunction that represents a reasonable approximation to ground state and is usually variationally determined.

A central quantity in both variational and projective QMC algorithms is the \emph{local energy}
\begin{equation}
\label{eq:le}
    e_L(x) = { \langle x | H | \psi_T \rangle \over  \langle x  | \psi_T \rangle } =  { \sum_{x'} H_{x',x} \langle x' |  \psi_T \rangle \over  \langle x  | \psi_T \rangle }.
\end{equation}
In conventional QMC methods, the local energy can be computed efficiently to additive precision as long as the number of states $x'$ such that the Hamiltonian matrix elements $|H_{x,x'}| \neq 0$, at fixed $x$  is polynomial in the number of spins. This requirement is commonly realized for virtually all physically relevant Hamiltonians, for example for k-local ones.
In this setting, computing the ratio of overlaps in Eq.~\ref{eq:le} does not pose any complication since the trial wave function's amplitudes $\psi_T(x)$, as well as their logarithm  $\log(\psi_T(x))$ can be computed with essentially arbitrary numerical precision. As a consequence, the ratios $\langle x' |  \psi_T \rangle \over  \langle x  | \psi_T \rangle$ can also be efficiently evaluated with additive precision.
In addition to that, in conventional QMC algorithm, the \emph{zero-variance} property holds, namely, if $\psi_T$ is an eigenstate of the hamiltonian, then $e_L(x)$ is constant over the Hilbert space. Thus, if $\psi_T$ is the exact ground state $\psi_0$, then a single evaluation of $e_L(x)$ gives the exact ground state energy $E_0$, with no statistical fluctuations.

Following the QC-QMC approach of Ref~\cite{huggins2022unbiasing} the trial state
is to be prepared instead by a quantum circuit, and the overlaps $\psi_T(x) = \langle x| \psi_T \rangle $ are to be evaluated from quantum measurements. Contrary to the classical case, thus evaluating wave functions amplitudes on a quantum computer is an operation intrinsically subject to statistical noise. These overlaps can be estimated with additive accuracy, for example using the Hadamard test, as proposed in Ref~\cite{huggins2022unbiasing}.
While evaluating the amplitudes to additive precision is efficiently realizable on a quantum computer,  obtaining constant relative precision is typically much more demanding, as also discussed in Supplementary Section F of Ref. \cite{huggins2022unbiasing}. We notice now that the estimation of the local energy involves ratios of amplitudes and, in this sense, it is crucial to estimate the required number of measurements necessary to estimate the local energy with given additive precision. Almost by definition, trial correlated wave functions of interest for the QC-QMC are such that their amplitudes are not concentrated on a poly-sized subset of states $x$, otherwise they could be efficiently simulated by classical computers \cite{schwarz_simulating_2013}.
Because of this feature, correlated states are such that typical amplitudes must be, for normalization reasons, exponentially vanishing: $|\psi_T(x)|^2 \simeq 2^{-L}$. As a consequence, while additive precision in the estimates of $|\psi_T(x)|$ is possible, the relative precision in the estimation of $e_L(x)$, $\delta e /e_L (x)$, is controlled by $\delta \psi_T  /|\psi_T(x)| \sim 2^{L}$, the relative error in the overlap estimation.  This issue is a fundamental limiting factor of the QC-QMC method, an issue as severe as the sign problem in classical QMC, also mathematically stemming from a similar propagation of errors in exponentially small ratios \cite{troyer_computational_2005}.

In order to substantiate this general argument more quantitatively, we now emulate the computation of Eq.~\ref{eq:le} in the ideal quantum setting, i.e. hardware noise free conditions.
We consider the realistic case, where the compact trial state that we can prepare, $\psi_T$, is not exactly the ground state. For this purpose, we consider a simple Jastrow wavefunction
\begin{equation}
\label{eq:jas}
    \psi_T (x) = \exp\left( \lambda_1 \sum_{k}^L s^z_{k} s^z_{k+1} + s_2 \sum_{k}^L s^z_{k} s^z_{k+2} \right)
\end{equation}
with fixed parameters $\lambda_1=0.233$, $\lambda_2=0.083$. This trial state yields variational energies within a $1\%$ error for all system sizes.
Since both the trial and the ground states are positive definite, the overlaps $ \langle x  | \psi_T \rangle$ can be computed directly from the measurements distribution, taking the square root of the normalized frequency counts. Notice that this is the best case scenario for the quantum setting, as it does not require the Hadamard test method discussed in Ref. \cite{huggins2022unbiasing} to reconstruct the additional phase of the amplitudes.

\begin{figure*}[hbt!]
\includegraphics[width=1\textwidth]{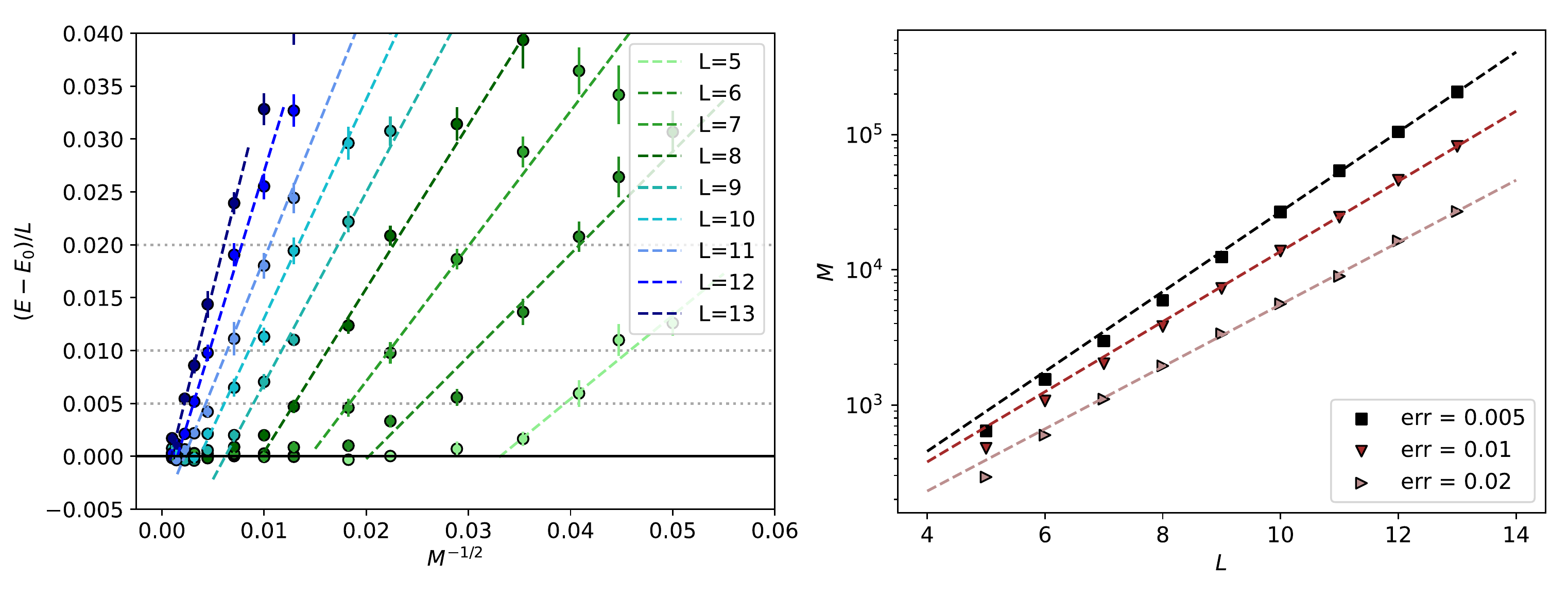}
\caption{
\emph{Left.} Error in the GFMC energy per site (compared to $E_0$) as a function of $M^{-1/2}$ for different system sizes. The trial wavefunction in this case is the Jastrow ansatz of Eq.~\ref{eq:jas}. Error bars obtained considering 16 statistically independent re-runs of the whole GFMC algorithm. Dashed lines are fits valid in the linear regime (see text). The linear fits are used to better localize the crossing with constant error lines at 0.005, 0.01, and 0.02 (dotted lines). \emph{Right.} Number of shots $M$ required to reach target errors in the energy per site. Dashed lines are exponential fit $M=a 2^{bL}$ to data points with $L>6$. The parameters found are respectively $(a=29.9, b =0.982)$ for an error of 0.005, $(34.7,0.862 )$ for an error of 0.01, and $(27.7, 0.764)$ for 0.02.
}
\label{fig:gf}
\end{figure*}

We emulate 16 independent measurements processes,  for different systems sizes, from $L=6$ to $12$ qubits. The overlap is estimated using a finite number of shots $M$.
In Fig.~\ref{fig:le} we plot the local energy for all the Hilbert space configuration $x$ (sorted by decreasing overlap magnitude), as well as the effective ``noisy" overlaps.
Crucially, we observe that the number of shots $M$ needs to scale exponentially with the system size to keep a fixed additive error for all system sizes.

The noisy evaluation of the local energy
also impacts the whole imaginary time projection, as it can be seen analyzing one of the simplest projective QMC algorithms, known as the Green's function Monte Carlo \cite{trivedi_ground-state_1990,runge_finite-size_1992,Becca2017}.
The discrete evolution of the walker, $x$ in imaginary time follows a stochastic process generated by the so-called importance sampling Green's function
\begin{align}
\label{eq:gfimportance}
    \bar{G}_{x,x'} &= \psi_T(x') G_{x,x'} / \psi_T(x) \\
    G_{x,x'} &= \langle x' | \Lambda 1 - H | x \rangle = \Lambda \delta_{x,x'} - H_{x,x'} \label{eq:gf}
\end{align}



The transition probability matrix used to update a walker is given by
\begin{align}
\label{eq:probgf}
    p_{x,x'} &= \bar{G}_{x,x'} / b_x \\
    b_x &= \sum_{x'}\bar{G}_{x,x'}  = \Lambda - e_L(x),
\end{align}
and also depends on the local energy.
Therefore, the effective Markov chain implemented by the QC-QMC scheme does not realize the correct imaginary time projection, as achieved in conventional GFMC.


To consider all these source of errors together, we estimate the GFMC energy by running the full algorithm, once again emulating the statistical error in the evaluation of the overlaps.
For each $(L,M)$ set of parameter we realize 16 independent Markov chains. The final energy is evaluated by reweighting the sequence of $e_L(x)$ generated during the trajectory (see Ref.~\cite{Becca2017} and Appendix \ref{app:gfmc} for details).
In Fig.~\ref{fig:gf} we plot the final result of the computed energy (per spin) as a function of the measurements budget $M$, for different system sizes.

The residual error scales as $c(L) M^{-1/2}$ except for small sizes $L$. In this case, for $M >> 2^L$, the error in the overlaps (which are precomputed and are unchanged during the random walk) translates into a small variational error that can be efficiently eliminated by the imaginary time projection.
The prefactor $c(L)$ increases with $L$, in such a way that
the number of measurements $M^*$ needed to achieve a fixed error, for instance of $0.005, 0.01$ and $0.02$ in units of energy per site, increses exponentially (cfn. Fig.~\ref{fig:gf}.)
For instance, to achieve an absolute error of $0.005$ in the energy per site, corresponding to a relative error of $0.4 \%$, we observe a scaling of $ M^* \approx 30 \times 2^{0.98~L}$. As a matter of comparison, this scaling is essentially on par with the worst-case scenario of performing exact diagonalization of the Hamiltonian, while achieving a relative error that is significantly larger than what can be obtained using classical variational approaches based on matrix product states or neural quantum states.

If we consider a $L=40$ spin system, marginally beyond what accessible with exact diagonalization on classical hardware, we estimate that the running-time requirements of the QC-QMC are extremely demanding, requiring more than $1.6 \times 10^{13}$ quantum measurements.
Assuming a best case scenario where an accurate trial state can be prepared with linear scaling depth, i.e. 40 layers, and assuming a realistic logical rotation gate clock of 10 kHz\cite{fowler2018low}, then the total wall-time budget required to collect the samples is $6.6 \times 10^{10} s$, i.e. about 2100 QPU years, also neglecting qubit reset, measurement time, and communication latency.
Finally, we report that the specific choice of the trial state is not crucial. We repeat the assessment using the exact ground state as trial state, and we recover again the exponential cost of the QC-QMC simulations. (see Appendix Fig.~\ref{fig:gfgs}.)

To conclude, we have provided a detailed analysis of the running time of a recently proposed hybrid QC-QMC algorithm. By considering the behavior of its core routine on a prototypical correlated model, we have shown that the scaling of this approach is comparable to exact diagonalization methods, thus leaving little room for a potential quantum advantage.
Our analysis cannot exclude, by construction, the existence of a specific combination of model and trial wave function for which the best classical algorithm scales with a worse exponential behavior than QC-QMC. Potential strategies to reach this regime of quantum advantage have been suggested in Ref. \cite{huggins2022unbiasing}, when alluding to using ``more sophisticated wave function forms". However, this extension of QC-QMC is at present unrealized, and beyond the scope of what presented in \cite{huggins2022unbiasing}. Speculating on the nature of this extension, we can however remark already that a fundamental limitation is that the target correlated state must be sparsely representable in the basis of the ``more sophisticated" walkers. Under this circumstance, it is yet to be demonstrated whether QC-QMC offers any advantage over conventional variational quantum algorithms and quantum phase estimation itself.

In general, we hope that our analysis will help elucidate current challenges in the implementation of QMC-inspired hybrid projection algorithms on quantum computers, rather than entirely excluding their viability. In this sense, the efficient evaluation of local energies estimators on quantum computers constitutes a rather stringent constraint in the design of competitive hybrid quantum algorithms. We hope our work will stimulate further research at the interface between quantum computing and Monte Carlo methods where these issues can be circumvented.

\textbf{Acknowledgement.} We acknowledge useful discussions with Sandro Sorella, and Shiwei Zhang.
G.M. acknowledges the support of SNSF Professorship grant.

\textbf{Author contributions.}
G.M. and G.C. contributed equally to the design of the research. G.M. performed the numerical simulations.

\appendix

\section{Appendix A: GFMC using the exact trial state}

Here we show that our observations are general and not specific to choice of the Jastrow trial wavefunction, and therefore, whether the trial state is efficiently representable by a quantum circuit or not.
We put ourselves in the best case scenario where we assume that we can prepare the exact ground state as the trial state $| \psi_T \rangle =  |\psi_0 \rangle$.
Notice that, in the quantum setting, the zero variance property does not hold\cite{wecker2015progress}, therefore one would not be able to certify the successful preparation of the ground state only within VQE (assuming that the exact solution is unknown).
If the computation of the local energy would be efficient in the quantum setting, one would simply observe that $e_L(x)$ is constant for all $x$ in the computational space.
This is precisely what happens in the conventional GFMC method with $| \psi_T \rangle =  |\psi_0 \rangle$.

Even in this unrealistically good scenario the QC-QMC method delivers an exponential cost with the system size (see Appendix Fig.~\ref{fig:gfgs})

\begin{figure*}[hbt!]
\includegraphics[width=1\textwidth]{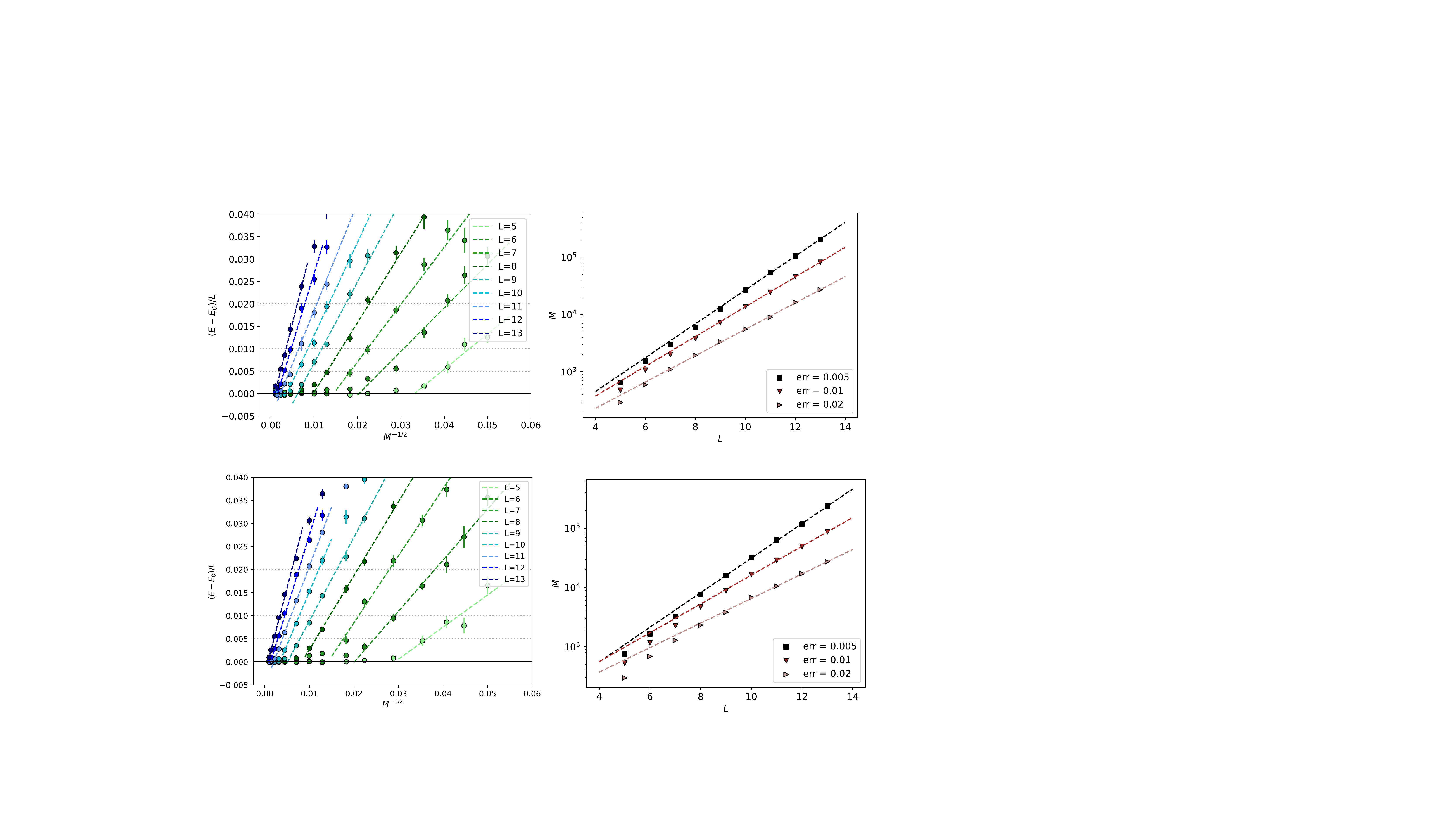}
\caption{
\emph{Left.} Error in the GFMC energy per site (compared to $E_0$) as a function of $M^{-1/2}$ for different system sizes. The trial wavefunction in this case is the exact ground state. Notice that in this particular case, the conventional GFMC would already give the exact energy with one single evaluation of the local energy.  Error bars obtained considering 16 statistically independent re-runs of the whole GFMC algorithm. Dashed lines are fits valid in the linear regime (see text). The linear fits are used to better localize the crossing with constant error lines at 0.005, 0.01, and 0.02 (dotted lines). \emph{Right.} Number of shots $M$ required to reach target errors in the energy per site.
Dashed lines are exponential fit $M=a 2^{bL}$ to data points with $L>6$. The parameters found are respectively $(37.8 , 0.970)$ for an error of 0.005, $(59.6,  0.809 )$ for an error of 0.01, and $(55.2 0.689)$ for 0.02.
}
\label{fig:gfgs}
\end{figure*}

\section{Appendix B: Details of GFMC algorithm}
\label{app:gfmc}

The GFMC algorithm defines a Markov process that yields, after a large number $n$ of iterations a probability distribution for the walker, the set $(w,x)$, which approximates the ground state wavefunction, namely
\begin{equation}
    \int dw w P_n(x,w) = \langle x | \psi_n \rangle \rightarrow \psi_0(x)
\end{equation}.

To do so, starting from an arbitrary position $x_i$, a new position $x_{i+1}$ is generated following Eq.~\ref{eq:probgf}, while formally the walker weight is updated via $w_{i+1} = w_{i} b_x$.
In practice, if the target observable is the energy, one simply needs to store only the sequence of $b_x$ generated along the chain.

Following Ref.~\cite{Becca2017} the ground state energy in single-walker GFMC is obtained from a single Markov chain of length $N$ by
\begin{equation}
\label{eq:gsfromgfmc}
    E_{GS} = { \sum_n^N G_n^l e_L(x_n) \over \sum_n^N G_n^l }
\end{equation}
where
\begin{equation}
    G_n^l = \prod_{i=1}^l b_x({n-i})
\end{equation}
In this case we set $l=100$, a value that ensures a sufficiently long imaginary time-projection and yields the correct ground state energy when the Markov chain is generated by a conventional GFMC.
In the main text, we generate 16 independent Markov chains, with different realization of quantum noise, so we obtain 16 different estimates of Eq.~\ref{eq:gsfromgfmc}. The error bars in Fig.~\ref{fig:gf} and ~\ref{fig:gfgs} are given by this sampling over the quantum noise realizations.

We notice that also Eq.~\ref{eq:gsfromgfmc} features the ratio of two stochastic quantities, but this does not produce numerical instabilities in the conventional method, provided a reasonable choice of $l$, as the numerator and denominator are highly correlated.

\begin{figure}[hbt!]
\includegraphics[width=1\columnwidth]{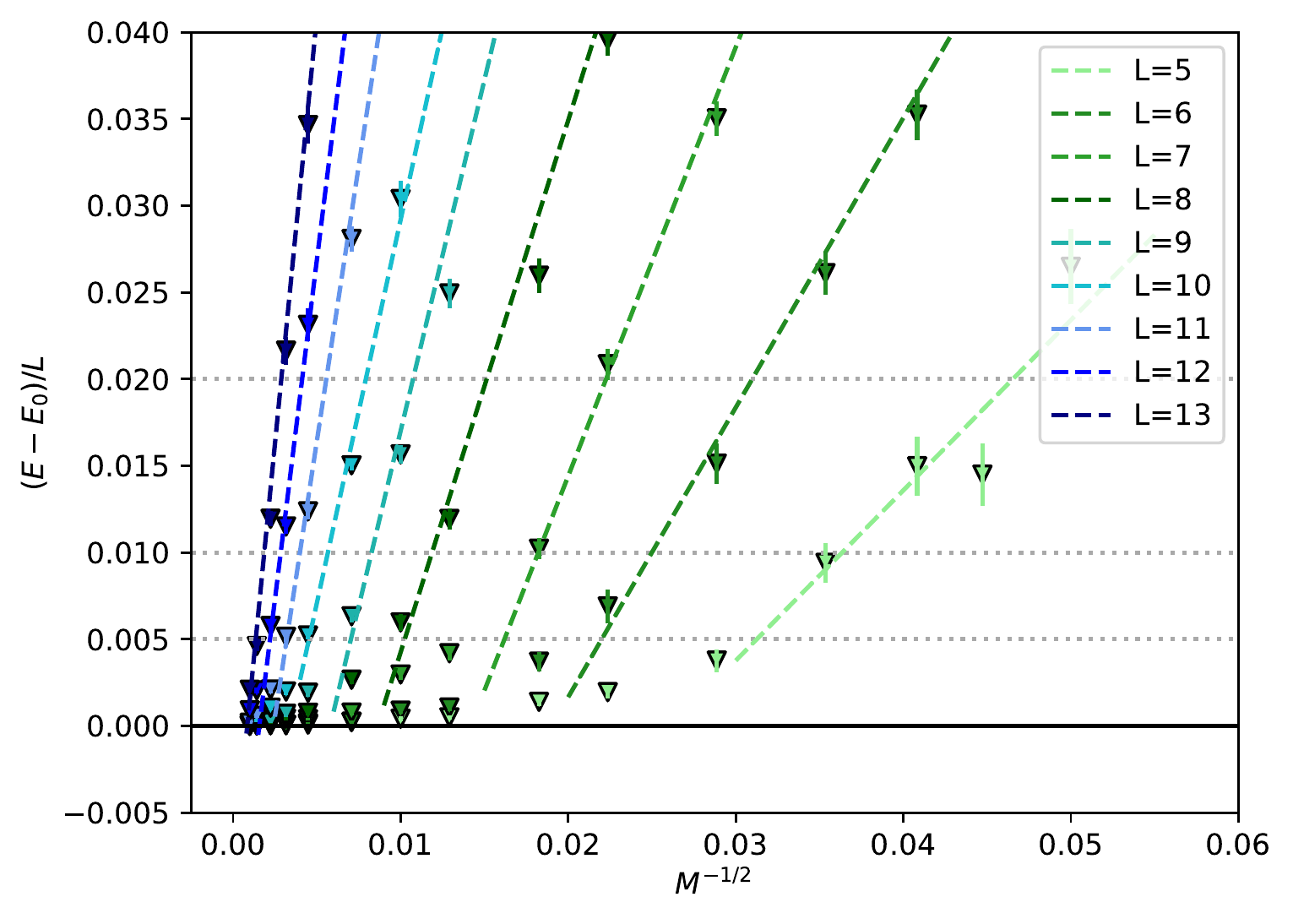}
\caption{
\emph{Left.} Error in the average local energy per site (compared to $E_0$) as a function of $M^{-1/2}$ for different system sizes. The trial wavefunction in this case is the exact ground state. Notice that in this particular case, the conventional method would already give the exact energy with one single evaluation of the local energy.  Error bars are obtained considering 16 statistically independent re-runs of the whole GFMC algorithm.
}
\label{fig:aveengs}
\end{figure}

We  numerically prove that this ratio is not the source of the exponential scaling observed in Fig.~\ref{fig:gf} and \ref{fig:gfgs}, even in presence of quantum measurement noise.
To do so, we repeat the analysis of Fig~\ref{fig:gfgs}, but we plot the average local energy value (Fig~\ref{fig:aveengs}), sampled along the Markov chain, without the reweighting of  Eq.~\ref{eq:gsfromgfmc}.

Even tough we use the exact ground state energy as  trial wavefunction, we observe the same striking $M$ dependence of the data, as in Fig.~\ref{fig:gfgs}.
This shows that the overall exponential scaling of the resources of the QC-QMC method is not due to the particular choice of our single walker GFMC, but it is a general manifestation of the uncontrolled statistal error of the local energy, a quantity which is central in every QMC algorithm.

%

\end{document}